\documentclass[aps,prl,superscriptaddress,twocolumn,twoside,a4paper,floatfix,nofootinbib]{revtex4-1}
\usepackage{hyperref,graphicx,amsmath,amssymb,mathpazo,color}
\usepackage[normalem]{ulem}

\usepackage{amsfonts}
\usepackage{latexsym}
\usepackage{amsmath}
\usepackage{amssymb}
\usepackage{amsfonts}
\usepackage{amsthm}
\usepackage{mathrsfs}
\usepackage{natbib}
\usepackage{bbm}
\usepackage{color,verbatim}
\DeclareMathAlphabet{\mathrsfs}{U}{rsfs}{m}{n}
\DeclareMathAlphabet{\mathpzc}{OT1}{pzc}{m}{it}
\DeclareMathAlphabet{\matheus}{U}{eus}{m}{n}
\DeclareMathAlphabet{\mathbbold}{U}{bbold}{m}{n}

\newcommand{\ba}{\begin{eqnarray}}
\newcommand{\be}{\begin{equation}}
\newcommand{\ee}{\end{equation}}

\newcommand{\ea}{\end{eqnarray}}
\newcommand{\ban}{\begin{eqnarray*}}
\newcommand{\ean}{\end{eqnarray*}}

\newcommand{\ket}[1]{|#1\rangle}
\newcommand{\bra}[1]{\langle#1|}

\begin{document}
\title{Entanglement enhances cooling in microscopic quantum fridges}
\author{Nicolas Brunner}
\affiliation{D\'epartement de Physique Th\'eorique, Universit\'e de Gen\`eve, 1211 Gen\`eve, Switzerland}
\affiliation{H. H. Wills Physics Laboratory, University of Bristol$\text{,}$ Tyndall Avenue, Bristol, BS8 1TL, United Kingdom}
\author{Marcus Huber}
\affiliation{Department of Mathematics, University of Bristol, University Walk, Bristol BS8 1TW, United Kingdom }
\affiliation{ICFO-Institut de Ciencies Fotoniques, Mediterranean Technology Park, 08860 Castelldefels, Barcelona, Spain}
\affiliation{Universitat Autonoma de Barcelona, 08193 Bellaterra, Barcelona, Spain}
\author{Noah Linden}
\affiliation{Department of Mathematics, University of Bristol, University Walk, Bristol BS8 1TW, United Kingdom }
\author{Sandu Popescu}
\affiliation{H. H. Wills Physics Laboratory, University of Bristol$\text{,}$ Tyndall Avenue, Bristol, BS8 1TL, United Kingdom}
\author{Ralph Silva}
\affiliation{H. H. Wills Physics Laboratory, University of Bristol$\text{,}$ Tyndall Avenue, Bristol, BS8 1TL, United Kingdom}
\author{Paul Skrzypczyk}
\affiliation{ICFO-Institut de Ciencies Fotoniques, Mediterranean Technology Park, 08860 Castelldefels, Barcelona, Spain}

\begin{abstract}
Small self-contained quantum thermal machines function without external source of work or control, but using only incoherent interactions with thermal baths. Here we investigate the role of entanglement in a small self-contained quantum refrigerator. We first show that entanglement is detrimental as far as efficiency is concerned---fridges operating at efficiencies close to the Carnot limit do not feature any entanglement. Moving away from the Carnot regime, we show that entanglement can enhance cooling and energy transport. Hence a truly quantum refrigerator can outperform a classical one. Furthermore, the amount of entanglement alone quantifies the enhancement in cooling. 
\end{abstract}

\maketitle

\section{Introduction}

The study of quantum thermal machines has a long history, from the thermodynamic analysis of lasers \cite{Ram56,ScoSch59,GeuSchSco67}, to considerable work on quantum cycles and the second law \cite{GevKos96,PalKosGor01,BenBroMei02,LinChe03,Scully03,HumLin05,SegNit06,HenMicMah06,HenMahMic07,BouTan07,QuaLiuSun07,AllHovMah10,Scully11,lutz}. Recently, models of small self-contained quantum thermal machines \cite{LinPopSkr10,SkrBruLin11,BLPS,CorrPalAde12,LevKos12} have attracted attention. The key feature of such machines is that they function without any external source of work or control. Only incoherent interaction with thermal baths are required. Interestingly, there exist no fundamental limit on the size of such machines \cite{LinPopSkr10}, nor on their efficiency \cite{SkrBruLin11}. Their main interest resides in their simplicity, which makes them an ideal test-bed for exploring quantum thermodynamics. 

Beyond the fundamental interest, these models may also turn out to be relevant from a more practical point of view. Proposals for experimental realizations were made, considering various physical platforms. A scheme using superconducting qubits (3 flux-biased qubits) driven by current noise was discussed in Ref. \cite{chenli}, while an electronic setup based on quantum dots was presented in \cite{giovanetti} (see also \cite{ButSot12}). Finally, Ref. \cite{mari} discussed the cooling of an optomechanical systems.

From a more fundamental point of view, an important question is whether quantum effects play any significant role in small self-contained thermal machines.
Indeed, although these machines are described within the quantum formalism, it is not immediately clear to what extent their working is inherently quantum. 
One can give an heuristic account of the functioning of the machine in classical terms.

Here, our aim is to establish the role of quantum effects in self-contained quantum thermal machines. Our main focus will be on the concept of entanglement, the paradigmatical quantum effect. Hence, if entanglement turns out to play a role in self-contained quantum thermal machines, this would make it clear that the working of such machines is truly quantum mechanical. Moreover, it would then raise the question of whether entanglement can enhance the performance of such machines. 

Below, we address these questions focusing on the model of the smallest possible self-contained quantum refrigerator \cite{LinPopSkr10,SkrBruLin11}. We first show that in the regime of high efficiency, that is machines operating with efficiency close to Carnot limit, the machine does not feature any entanglement. Hence, entanglement appears to be \emph{detrimental} as far as efficiency is concerned since an entangled state cannot get close to Carnot efficiency. Next, moving away from the high efficiency regime, we show that there exist regimes featuring entanglement. In fact, a wide variety of types of entanglement can be found in our system---including genuine multipartite entanglement---depending upon the external conditions. Finally, we show that this entanglement is useful, as it enhances cooling and energy transport. Specifically, given an object to cool and a set of resources (for instance fixing the temperatures of the heat baths), we show that a refrigerator featuring entanglement can outperform a 'classical' refrigerator (i.e. featuring no entanglement), as it allows to cool the object to lower temperatures. Moreover, we demonstrate that the improvement grows monotonically with entanglement measures, strongly suggesting of a functional relationship. 

We emphasize that the above results are not general, but refer to the specific fridge model discussed in Refs \cite{LinPopSkr10,SkrBruLin11}. However, as this model represents the simplest possible refrigerator, it is in some sense the most fundamental one. Results obtained here are therefore bound to reflect some basic facts about quantum thermal machines; in a way or another, they are likely to have implications for all machines.

\section{Quantum fridge model} 
\begin{figure}
  \includegraphics[width=0.9\columnwidth]{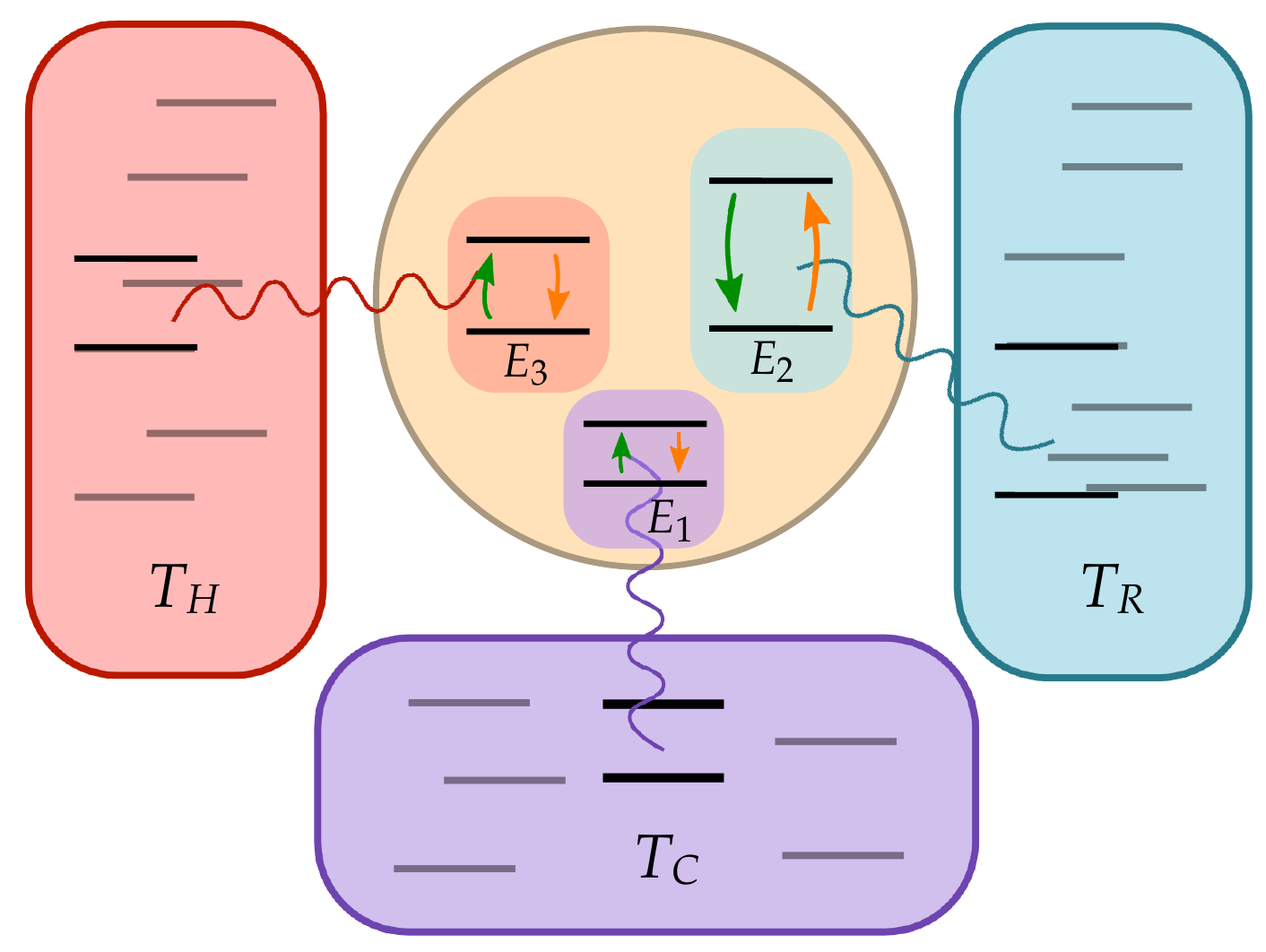}
  \caption{Schematic diagram of the quantum refrigerator. The fridge consists of three qubits (inside the yellow circle), each in weak thermal contact (wiggly lines) with a bath at a different temperature. The qubits interact via the weak interaction Hamiltonian $H_{int}$, which couples the two degenerate levels $\ket{010}$ and $\ket{101}$, depicted by the arrows. The lower qubit (purple) is the object to be cooled. At equilibrium, it reaches a temperature $T_S<T_C$. The other two qubits (red and blue) are the machine qubits, connected to heat baths at temperatures $T_R$ and $T_H$.}
\label{schematic}
\end{figure}

We start by briefly reviewing the model of the smallest quantum refrigerator of Ref.~\cite{LinPopSkr10,SkrBruLin11} (see Fig.~\ref{schematic}), which we will focus on in this work. Let us consider three qubits, which in the absence of interaction have vanishing ground state energies and excited state energies $E_j$ $(j=1,2,3)$. The free Hamiltonian of the refrigerator is thus given by 
\ba H_0= E_1 \Pi_1+E_2 \Pi_2+E_3 \Pi_3 \ea
where $\ket{1}_j$ is the exited state of qubit $j$, (i.e. energy eigenstate at energy $E_j$), and $\Pi_j= \ket{1}_j\bra{1}$ the corresponding projector. 
We choose the energy levels such that $E_2 = E_1 + E_3$ and $E_1 \neq E_3$. Given this condition the Hamiltonian contains only two states degenerate in energy, $\ket{010}$ and $\ket{101}$.
Next we consider weakly coupling the qubits by placing an interaction between them. In the weak coupling regime, the main effect of a generic 3-qubit Hamiltonian is to couple only states degenerate in energy -- in this case $\ket{010}$ and $\ket{101}$\cite{footnote1}; couplings between other states are suppressed by a factor inversely proportional to their energy difference. Thus, we focus here on this main effect, i.e. the interaction Hamiltonian
\ba H_{int} = g \left(\ket{010}\bra{101} + \ket{101} \bra{010}\right). \ea
We impose $g \ll E_j$, which signifies the weak coupling regime, and also ensures that the eigenvalues and eigenstates of the fridge remain governed by $H_0$.

Finally, each qubit is taken to be in contact with a separate thermal reservoir. The temperatures of the reservoirs are denoted by $T_C$ (cold), $T_R$ (room), and $T_H$ (hot), for qubits 1, 2 and 3 respectively. The thermal contact between each qubit and bath is governed by Linbladian dissipative dynamics, which we model here using a simple reset model, the justification of which we shall comment on briefly. In this model, with probability $p_i \delta t$ per time $\delta t$, qubit $i$ is reset to the thermal state $\tau_i$, at the temperature of its bath, while for all other times it evolves unitarily according to the combined Hamiltonian $H_0 + H_{int}$. That is, in this model thermalization events are taken to be rare but strong events. The equation of motion for the refrigerator using this model of dissipation \cite{LinPopSkr10} is given by the Master equation 
\ba\label{Master} \frac{\partial \rho}{\partial t} = -i [H_0 +H_{int}, \rho] + \sum_{i} p_i (\tau_i \otimes Tr_i(\rho) - \rho) \ea
where $\tau_i = r_i \ket{0}_i\bra{0} + (1-r_i) \ket{1}_i\bra{1}$ with $r_i= 1/(1+e^{-E_i/T_i})$. In general one would expect there to be additional terms in equation \eqref{Master}, corresponding to dissipative dynamics on qubit $j$ originating from the combination of the interaction Hamiltonian and the dissipative dynamics on qubit $i\neq j$. In other words, one may expect each qubit to be effectively in contact with all three baths due to the interaction Hamiltonian \cite{Tal86,CorrPalAde12}. However, in the regime where $p_i \approx g \ll E_i$, these additional effects, whose strength is approximately $gp_i \ll g$ can be safely neglected. 

Here our focus is on the stationary state (i.e. long term behaviour) of the refrigerator, $\rho_S$, which satisfies $\dot{\rho}_S=0$ i.e.
\ba i [H_0 +H_{int}, \rho_S] = \sum_{i} p_i (\tau_i \otimes Tr_i(\rho_S) - \rho_S). \ea

As shown in \cite{SkrBruLin11}, this equation can be solved analytically for all values of the parameters. The solution takes the form
\ba \label{rhoS} \rho_S = \tau_1 \tau_2 \tau_3 + \gamma \sigma \ea
where $\gamma$ is a dimensionless parameter depending upon all parameters of the model (namely $p_i$, $g$, $E_i$, and temperatures $T_{C,R,H}$), and $\sigma$ is a traceless matrix with a single off-diagonal term (see \cite{SkrBruLin11} for details). The important property of the solution is that it can be shown that the refrigerator cools qubit 1 whenever $\gamma>0$. In this case, one finds that qubit 1 is in a stationary state that is diagonal, with corresponding temperature $T^S<T_C$. Moreover, the efficiency of the refrigerator tends to the Carnot limit in the limit $\gamma \rightarrow 0$.

\section{Around the Carnot point} 
Let us first discuss the properties of $\rho_S$ for those refrigerators which are operating close to the Carnot efficiency -- which from hereon we refer to as refrigerators around the \emph{Carnot point}. From inspection of Eq.~\eqref{rhoS}, it is clear that for $\gamma=0$, $\rho_S$ is a fully separable state, as it is nothing other than the direct product of thermal state for each qubit. Hence, no entanglement is present at the Carnot point. More interestingly, this statement remains true for a small region within the set of all $\rho_S$ in the vicinity of the Carnot point. Thus all refrigerators which are highly efficient function without entanglement. To see this let us first rewrite $\rho_S$ in the following form 
\ba \label{rhoS2} \rho_S = w \ket{GHZ}\bra{GHZ} + (1-w) \sigma_{diag} \ea
where $\ket{GHZ} = ( \ket{010} +  i\ket{101})/\sqrt{2}$ is tripartite entangled state (of the Greenberger-Horne-Zeilinger form), and $\sigma_{diag}$ is a diagonal density matrix, hence corresponding to a fully separable state. While there is no unique notion of entanglement in multipartite systems, it turns out that the entanglement of states of the form \eqref{rhoS2} can be conveniently characterized. 

In the vicinity of any Carnot point, the state $\rho_S$ has full rank and off-diagonal terms which are small compared to diagonal ones. Hence, in this regime, the state can be decomposed as $\rho_S=(1-\epsilon)\sigma'_{diag} +\epsilon\rho(p)$, where $\sigma'_{diag} $ is a diagonal separable state and $\rho(p)=p|GHZ)\rangle\langle GHZ|+(1-p)\frac{1}{8}\mathbbm{1}$. 
Since $\rho(p)$ is fully separable for $p\leq \frac{3}{11}$ \cite{Siewert}, it follows that $\rho_S$ is fully separable in the vicinity of any Carnot point. In fact, it can even be shown that any $\rho_S$ at the Carnot point has a small ball of fully separable state in the full Hilbert space around it---see Appendix A for more details.

\begin{figure*}{t!}
  \includegraphics[width=1.85\columnwidth]{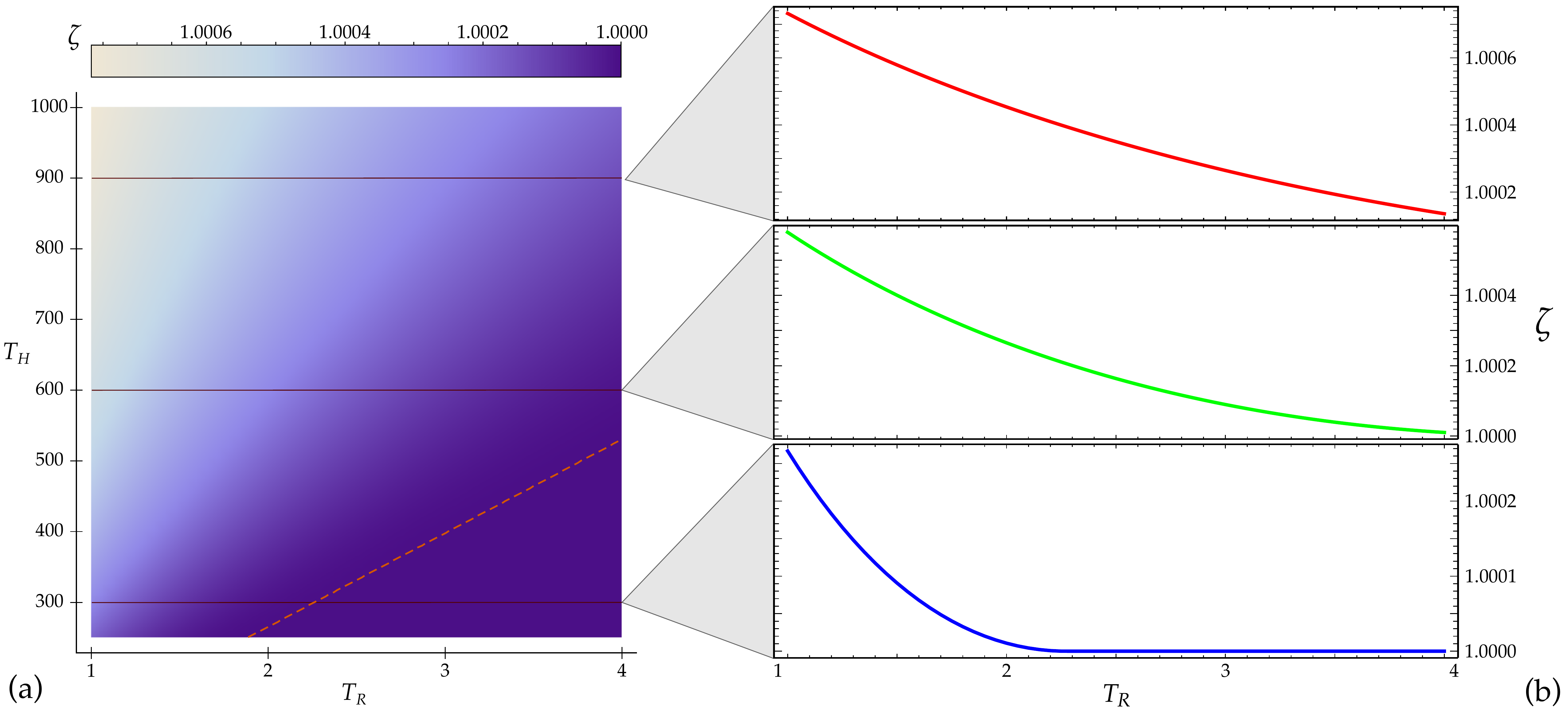}
  \caption{Entanglement can enhance cooling. (a) Density plot depicting the relative advantage $\zeta $ in cooling a qubit at $T_C=1K$, as a function of the available bath temperatures ($T_R$ and $T_H$), for optimal refrigerators. Entanglement provides an advantage ($\zeta>1$) for all points above the red dashed line. In the dark blue region, below the red dashed line, the optimal fridge is separable on all bipartitions ($\zeta=1$). For $T_R$ close to $T_C$ and sufficiently high $T_H$, entanglement is a widely present feature of the optimal quantum refrigerator. (b) Slices through the density plot for three different values of $T_H$, showing in more detail the relative advantage in cooling. The parameters used here are $p_C = 10^{-5}$, $E_1 = 1J$, and parameter bounds $p_R$,$p_H$,$g \le 10^{-4}$.}
\label{graphs}
\end{figure*}

\section{Entangled regimes} 
Next we ask whether there exists regimes in which entanglement is present in $\rho_S$. At this point it is useful to recall that entanglement can appear under several forms in a state of 3 qubits. Indeed, there can be bipartite entanglement along a given bipartition (e.g. qubit 1 versus qubits 2 and 3), or genuine tripartite entanglement. Here our main tool will be a class of entanglement witnesses developed in \cite{SeevinckGuhne,HMGH1} which allow one to fully characterize the entanglement of states of the form $\rho_S$. Moreover, these witnesses also provide a meaningful entropy based measure of multipartite entanglement \cite{Maetal} and necessary and sufficient conditions for biseparability for our system \cite{rafsanjani}. Formally, these witnesses are given by inequalities of the form
\ba W_{\mathcal{S}}(\rho)=\label{witness} 2 \left( |\rho_{3,6}|- \sum_{k\in\mathcal{S}}\sqrt{\rho_{k,k} \rho_{9-k,9-k}} \right) \leq 0 \ea
where $\rho_{i,j}$ denotes elements of the density matrix and the set $\mathcal{S}$ depends on the partition and type of entanglement one is interested in. When inequality \eqref{witness} is violated, its left-hand side gives the concurrence \cite{wootters} of $C|RH$ ($\mathcal{S}=\{2\}$), $R|CH$ ($\mathcal{S}=\{1\}$), $CR|H$ ($\mathcal{S}=\{3\}$) or the genuine multipartite concurrence (see Refs.~\cite{Maetal,Wuetal}) for $\mathcal{S}=\{1,2,3\}$. When inequality \eqref{witness} holds, no entanglement is present on the given bipartition.

Moving away from the Carnot point we find, by sweeping through the parameter space numerically, that there exists regimes where entanglement is present. In fact, most types of entanglement can be found. We find regimes where there is entanglement (i) along only a single bipartition of the system, (ii) on all three bipartitions at the same time, and (iii) genuine tripartite entanglement, the strongest form of multipartite entanglement. In Table~1 we characterize various entanglement regimes, giving the corresponding parameters of the refrigerator. Note that the only type of entanglement we did not observe is the following: entanglement on the $C|RH$ partition but no entanglement across the other two bipartitions.

\section{Entanglement enhances cooling} 
In the remainder of the paper, we investigate the usefulness of this entanglement that we have just uncovered in the fridge. 
We will see that entanglement can in fact enhance the performance of the refrigerator. For this we consider the task of cooling a qubit with given energy $E_1$, immersed in a bath at a given temperature $T_C$ with fixed coupling $p_1$. As a source of free energy, we have at our disposal two heat baths, at temperatures $T_R$ and $T_H$ (again assuming that $T_C<T_R<T_H$). The challenge is then to adjust the remaining parameters in order to minimize the temperature of the qubit in its stationary state. The free parameters are the energy of the hot qubit $E_3$, the thermalization coefficients for the machine qubits $p_2$ and $p_3$, and the interaction strength $g$. Indeed some of these parameters constrained by the weak coupling assumption. We require that $g \ll E_i$, $p_j \ll E_i$, $p_j g \ll g$ and $p_j g \ll p_j$, which can be enforced by choosing a cutoff for $p_j$ and $g$. We observe that all of our conclusions below remain valid independent of the precise choice of this cutoff. The only change is that the strength of the effect becomes weaker as we make the constraints stronger, as is intuitively expected. We comment further on this at the end of this section.

First, considering all possible fridges, we look for the one achieving the best cooling, i.e. the smallest value of $T_S$. Next, we repeat this optimization, but now adding the constraint that no entanglement is present in the fridge. More precisely, we find the optimal cooling (now denoted $T_S^*$), imposing that the stationary state $\rho_S$ satisfies all the entanglement witness inequalities \eqref{witness} (and their relevant symmetries), hence ensuring separability across every bipartition. The results are presented in Fig.~\eqref{graphs}. We observe that the cold qubit can be cooled to lower temperatures when no restrictions are placed, compared to the case when the system is constrained to be separable. In the regime where $T_R \ll T_H$, entanglement provides an enhancement in cooling, which is quantified by the ratio 
\ba \zeta = \frac{T_C-T_S}{T_C-T_S^*}. \ea 
When restricting to optimal machines, we find that the entanglement is almost always preferentially between the room qubit and the other two. Only if $T_R\approx T_C$ and $T_R\ll T_H$ does entanglement between the other two bipartitions also appear. An intriguing aspect of this behaviour is that the entanglement thus appears to be preferentially between the partitions \emph{energy in vs energy out}, since in the stationary state it is the room qubit which is heated up, and the cold and hot qubits which are cooled down. Hence, it seems that quantum coherence (here entanglement) enhances energy transport, a phenomenon that has received considerable attention in the context of photosynthetic complexes \cite{Engel}. 

It is important to note that, although the cooling enhancement observed here is relatively small, it is nevertheless much larger than the uncertainty in the result (which is $\mathcal{O}(gp)$) and hence represents a genuine effect, and not a mere consequence of our approximate dynamics. Not unexpectedly, the optimal cooling is obtained by maximizing the interaction coupling as well as the thermalization rates. Here these parameters must be constrained in the optimization in order to remain within the regime of validity of the master equation \eqref{Master}. It would be interesting to consider more general master equations (see e.g. \cite{CorrPalAde12}), accounting for more sophisticated thermalization processes. In this case, one could describe machines working at much higher rates, in which entanglement may become even more beneficial. We leave it for future research to explore this direction.

\begin{figure}
  \includegraphics[width=\columnwidth]{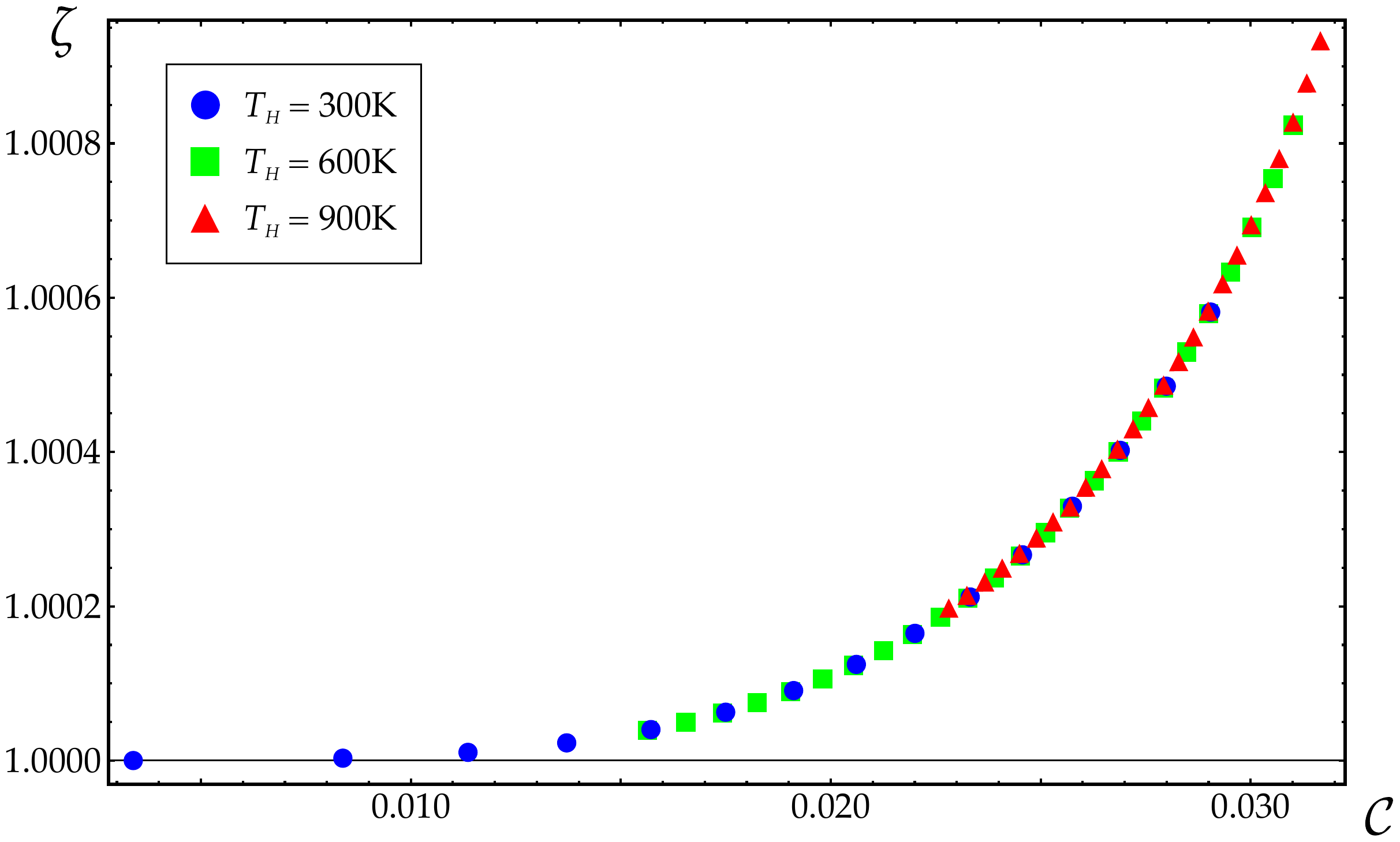}
 \caption{Cooling advantage is determined by the amount of entanglement. Plot of relative cooling enhancement $\zeta$ against amount of entanglement on the bipartition $R|CH$ (measured by the concurrence $C$), as evaluated from Fig.2a. Since all points lie on a single curve, it follows that $\zeta$ is determined solely by $C$, and does not depend on the temperatures of the baths $T_R$ and $T_H$. The fact that the behaviour is monotonic strongly suggests a functional relationship. For convenience the data plotted corresponds to the three horizontal slices of Fig.~2 (a). Taking random sample points from Fig.~2 (a) leads to a similar result.}
\label{concurrence}
\end{figure}

Finally, we investigate the link between the amount of entanglement on the bipartition $R|CH$ (as measured by the concurrence $\mathcal{C}$) and the relative cooling enhancement $\zeta$. 
Remarkably, $\zeta$ appears to be solely determined by the concurrence (see Fig.~\ref{concurrence}), and hence does not depend on the temperatures of the heat baths $T_R$ and $T_H$. This result strongly suggests a functional relationship between the relative cooling enhancement and concurrence. Indeed, the monotonic relationship observed shows that the more entanglement that is present, the larger the advantage one gains for cooling a system.

\begin{table*}
	\caption{\label{t:example} {\bf Regimes of the quantum refrigerator featuring entanglement.} Here we give the full list of parameters required to exhibit the different entanglement regimes present in the quantum refrigerator. The first row gives a point where there is genuine multipartite entanglement between all three qubits, whilst the second and third rows are points where there is only bipartite entanglement along a single bipartition. The last row shows a regime with bipartite entanglement across every bipartition but without genuine tripartite entanglement.} 
	\begin{ruledtabular}
	\begin{tabular}{|llll|lll|ll|l|lll|}
	$\mathcal{C}_{C|RH}$ & $\mathcal{C}_{R|CH}$ & $\mathcal{C}_{H|CR}$ & $\mathcal{C}_{CRH}$ & $T_C$ & $T_R$ & $T_H$ & $E_1$ & $E_3$ & $g$ & $p_1$ & $p_2$ & $p_3$ \\ \hline
	0.003 & 0.004 & 0.004 & 0.003 & 1.0 & 1.1 & $1.0 \times 10^{4}$ & 2.0 & 300 & $1.0 \times 10^{-4}$ & $1.0 \times 10^{-5}$ & $1.0 \times 10^{-3}$ & $1.0 \times 10^{-5}$ \\ 
	0 & 0 & 0.00002 & 0 & 1.0& 2.0 & $1.0 \times 10^{4}$ & 4.4 & 379 & $3.4 \times 10^{-4}$ & $1.3 \times 10^{-5}$ & $1.6 \times 10^{-4}$ & $1.3 \times 10^{-5}$ \\  
	0 & 0.00005 & 0 & 0 & 1.0 & 42.6 & $1.0 \times 10^{4}$ & 4.4 & 421 & $2.8 \times 10^{-4}$ & $1.3 \times 10^{-5}$ & $8.3 \times 10^{-4}$ & $1.0 \times 10^{-5}$ \\  
	0.0008 & 0.005 & 0.004 & 0 & 1.0 & 1.1 & $1.0 \times 10^{4}$ & 2.0 & 300 & $1.0 \times 10^{-4}$ & $1.0 \times 10^{-5}$ & $2.0 \times 10^{-4}$ & $1.0 \times 10^{-5}$ \\
	\end{tabular}
	\end{ruledtabular}
\end{table*}

\section{Conclusions} 
We have discussed the role of entanglement in the smallest self-contained quantum refrigerator. Entanglement turns out to be a feature of a wide range of operation regimes, with the notable exception of the Carnot point and its vicinity. Crucially, this entanglement is not a mere byproduct, but its presence is beneficial: the fridge is able to cool to lower temperatures when it becomes entangled. Moreover, our results show that the extra quantum performance is directly related to measures of entanglement.

One important question is to what extent the results obtained here tell us about entanglement and its usefulness in general thermal machines. In \cite{BLPS} it was argued that any thermal machine which is operating close to the Carnot efficiency is necessarily functioning in the same manner as the machines studied here. This is precisely the regime where no entanglement is present, hence we can conclude that close to the Carnot efficiency no thermal machine is entangled (at least regarding those degrees of freedom directly relevant to the cooling process). What would be more interesting however is to go in the other direction, and make statements about the regime where entanglement is beneficial. At the present such general results appear to be beyond reach, but any such statements in this direction would represent significant progress.

Another question is whether entanglement can enhance the performance of machines producing work. The model discussed here, when analysed appropriately (in particular by replacing the cooled qubit by a `weight' which can store work), can also function as a small work-producing heat engine \cite{BLPS}. Hence an analysis of the presence and role of entanglement in this system will also be of particular interest, especially given the recent results \cite{work1,work2,work3,work4} where the role of entanglement in work extraction from quantum systems is explored. 

To conclude, we believe that the present results provide evidence that entanglement plays a significant role in thermodynamic processes. Clearly, turning these preliminary results into a general and quantitative understanding of the exact role played by entanglement in thermodynamics is highly desirable. 

\emph{Note added.} While finishing this work, we became aware of the work of Correa and colleagues \cite{CorrPalAde12} discussing the effect of quantum discord in the small self-contained quantum refrigerator. 

\emph{Acknowledgments.} We thank J. Reid for contributions in early stages of this work. We acknowledge financial support from the Swiss National Science Foundation (grant PP00P2\_138917), the FP7-MarieCurie grant 'Quacocos', the ERC (Advanced Grant ÒNLSTÓ), and the Templeton Foundation.

\section{Appendix A}

In this appendix we show that around any Carnot point (i.e. any any machine which is operating at the Carnot efficiency) there is a ball of states which are fully biseparable (i.e. separable on any bipartition). It is also shown that in some circumstances, there is a ball of fully separable states.

More precisely, the first statement that we would like to prove is that there exists an $\epsilon > 0$, such that for all states $\sigma$ which satisfy $\| \rho_S- \sigma \|_1 \le \epsilon$ are biseparable, where $\rho_S$ is a stationary state of the refrigerator corresponding to a machine at Carnot efficiency.

The first important point is that all Carnot points have the form $\rho_S = \tau_C \otimes \tau_R \otimes \tau_H$, where each $\tau_i$ is a thermal state at a strictly positive temperature $T_i > 0$ (since if we consider the cold bath to be at absolute zero there is no cooling to be achieved). As such, it is immediately clear that $\rho_S$ is a full rank fully separable state (in fact it is a direct product state). The fact that $\rho_S$ is full rank guarantees that there is a ball of states around it (i.e. it is strictly in the interior of the set of quantum states). This ball however may contain entangled states, and so we must show that this ball contains within it a small ball of separable states.

Any states $\sigma$ which are diagonal in the same basis as $\rho_S$ are clearly also fully separable, and therefore we shall focus on states which are not diagonal. Let us consider first states $\sigma$ which are still within the interior of the set of quantum states but contain only a single pair of off-diagonal elements, i.e. states of the form
\begin{equation}
	\sigma = \rho_D + \sigma_{\mathbf{xy}}\ket{\mathbf{x}}\bra{\mathbf{y}} + \sigma^*_{\mathbf{xy}}\ket{\mathbf{y}}\bra{\mathbf{x}}
\end{equation}
where $\rho_D$ is diagonal in the same basis as $\rho_S$, and $\mathbf{x} = (x_C, x_R, x_H)$ collectively specifies the state for the three qubits in the energy eigenbasis (and analogously for $\mathbf{y}$). States of this form are in the class of so called X-states. For such states, the three nonlinear witnesses $W_i$, for $i = 1,2,3$, given by
\ba  W_i(\rho) = 2 \left( |\rho_{3,6}|- \sqrt{\rho_{i,i} \rho_{9-i,9-i}} \right) \leq 0 \ea
provide necessary and sufficient criteria for biseparability. This follows from the results of \cite{rafsanjani} and \cite{Wuetal}, using which one can show, respectively, that the $W_i$ provide upper and lower bounds on the concurrence. 

Since the witnesses $W_i$ are convex and $\rho_S$ is full rank, it is the case that the $W_i(\rho_S) < 0$, i.e. the witness is strictly negative on the Carnot point. It follows therefore that there is an $\eta$ such that for all $|\sigma_{\mathbf{xy}}| < \eta$, the state $\sigma$ is still separable. In other words, if we restrict to the set of states with only a single off diagonal element, then there exists an $\epsilon$ such that for all $\| \sigma - \rho_S \|_1 \le \epsilon$ the state is biseparable. 

To finish the proof note that we can simply take mixtures of the above states with only a single off diagonal element to define a set of states with an arbitrary number of off diagonal elements. That is, we write
\begin{equation}
	\sigma = \sum_i p_i (\rho_D + \sigma^i_{\mathbf{x}_i\mathbf{y}_i}\ket{\mathbf{x}_i}\bra{\mathbf{y}_i} + \sigma^{i*}_{\mathbf{x}_i\mathbf{y}_i}\ket{\mathbf{y}_i}\bra{\mathbf{x}_i})
\end{equation}
where again $\rho_D$ is full rank and all $\sigma^i_{\mathbf{x}_i\mathbf{y}_i}$ are sufficiently small relative to elements of $\rho_D$. From the above it follows that there is an $\epsilon > 0$ such that each state in the decomposition is both biseparable and positive semi definite. This thus demonstrates that there is a ball of fully biseparable states in the full Hilbert space around every Carnot point.

Moving on, the second claim we wish to prove is that around any Carnot point such that $\text{Tr}(\rho_S^2)<\frac{19}{24}$, there is a ball of fully separable states. To do so we shall use the fact that there is a ball of fully separable states around the maximally mixed state \cite{braunstein}. In particular, in our case the size of this ball is $\epsilon_\mathrm{sep}= \sqrt{\frac{2}{3}}$ \cite{Gurvits}. Hence, full separability of $\rho_S$ is ensured when 
\begin{equation}
\| \rho_S - \frac{\openone}{8} \|_2 < \sqrt{\frac{2}{3}}
\end{equation}
which implies the existence of a ball of separable states around $\rho_S$ when $\mathrm{Tr} \rho_S^2 < \frac{19}{24}$.


\begin{thebibliography}{99}
\bibitem{footnote1} Note that this effect is exploited in the experimental proposals for implementing the fridge \cite{chenli,giovanetti} (see also \cite{mari}).
\end{thebibliography}
\end{document}